\documentstyle[12pt,aaspp4]{article}
\begin{document}

\newcommand{\up}[1]{\ifmmode^{\rm #1}\else$^{\rm #1}$\fi}
\newcommand{\zdot}{\makebox[0pt][l]{.}}
\newcommand{\upd}{\up{d}}
\newcommand{\uph}{\up{h}}
\newcommand{\upm}{\up{m}}
\newcommand{\ups}{\up{s}}
\newcommand{\arcd}{\ifmmode^{\circ}\else$^{\circ}$\fi}
\newcommand{\arcm}{\ifmmode{'}\else$'$\fi}
\newcommand{\arcs}{\ifmmode{''}\else$''$\fi}

\title{The Araucaria Project. The Distance to the Local Group 
Galaxy WLM from Cepheid Variables discovered in a Wide-Field Imaging Survey
\footnote{Based on  observations obtained with the 1.3~m
telescope at the Las Campanas Observatory.
}
}

\author{Grzegorz Pietrzy{\'n}ski}
\affil{Universidad de Concepci{\'o}n, Departamento de Fisica, Astronomy
Group,
Casilla 160-C,
Concepci{\'o}n, Chile}
\affil{Warsaw University Observatory, Al. Ujazdowskie 4,00-478, Warsaw,
Poland}
\authoremail{pietrzyn@hubble.cfm.udec.cl}
\author{Wolfgang Gieren}
\affil{Universidad de Concepci{\'o}n, Departamento de Fisica, Astronomy Group, 
Casilla 160-C, 
Concepci{\'o}n, Chile}
\authoremail{wgieren@astro-udec.cl}
\author{Andrzej Udalski}
\affil{Warsaw University Observatory, Aleje Ujazdowskie 4, PL-00-478,
Warsaw,Poland}
\authoremail{udalski@astrouw.edu.pl}
\author{Igor Soszy{\'n}ski}
\affil{Universidad de Concepci{\'o}n, Departamento de Fisica, Astronomy
Group, Casilla 160-C, Concepci{\'o}n, Chile}
\affil{Warsaw University Observatory, Aleje Ujazdowskie 4,
PL-00-478,Warsaw, Poland}
\author{Fabio Bresolin}
\affil{Institute for Astronomy, University of Hawaii at Manoa, 2680 Woodlawn 
Drive, 
Honolulu HI 96822, USA}
\authoremail{bresolin@ifa.hawaii.edu}
\author{Rolf-Peter Kudritzki}
\affil{Institute for Astronomy, University of Hawaii at Manoa, 2680 Woodlawn 
Drive, Honolulu HI 96822, USA}
\authoremail{kud@ifa.hawaii.edu}
\author{Alejandro Garcia}
\affil{Universidad de Concepci{\'o}n, Departamento de Fisica, Astronomy
Group,Casilla 160-C,
Concepci{\'o}n, Chile}
\authoremail{agarcia@astro-udec.cl}
\author{Dante Minniti}
\affil{Pontifica Universidad Cat{\'o}lica de Chile, Departamento de
Astronomia y Astrofisica, Casilla 306, Santiago 22, Chile}
\author{Ronald Mennickent}
\affil{Universidad de Concepci{\'o}n, Departamento de Fisica, Astronomy
Group, Casilla 160-C,
Concepci{\'o}n, Chile}
\authoremail{rmennick@astro-udec.cl}
\author{Olaf  Szewczyk}
\affil{Warsaw University Observatory, Aleje Ujazdowskie 4, PL-00-478,
Warsaw,Poland}
\authoremail{szewczyk@astrouw.edu.pl}
\author{Micha{\l} Szyma{\'n}ski}
\affil{Warsaw University Observatory, Aleje Ujazdowskie 4, PL-00-478,
Warsaw, Poland}
\authoremail{msz@astrouw.edu.pl}
\author{Marcin Kubiak}
\affil{Warsaw University Observatory, Aleje Ujazdowskie 4, PL-00-478,
Warsaw, Poland}
\authoremail{mk@astrouw.edu.pl}
\author{{\L}ukasz Wyrzykowski}
\affil{Warsaw University Observatory, Aleje Ujazdowskie 4, PL-00-478,
Warsaw,Poland}
\affil{Institute of Astronomy, University of Cambridge, Madingley Road, CB3 0HA, UK}
\authoremail{wyrzykow@astrouw.edu.pl}

\begin{abstract}
We have conducted an extensive wide-field imaging survey for Cepheid variables
in the Local Group irregular galaxy WLM. From data obtained on 101 nights,
we have discovered 60 Cepheids which
include 14 of the 15 Cepheid variables previously detected by Sandage and Carlson.
Down to a period of 3 days, our Cepheid survey in WLM should be practically
complete. Importantly, we have found for the first time a long-period Cepheid
(P=54.2 days) in this galaxy, alleviating the puzzle that WLM with its many
blue, massive stars does not contain Cepheids with periods longer than about 
10 days. Our data define tight period-luminosity relations in V, I and the
reddening-free Wesenheit magnitude ${\rm W}_{\rm I}$ which
are all extremely well fit by the corresponding slopes of the LMC Cepheid
PL relation, suggesting no change of the PL relation slope down to a
Cepheid metal abundance of about -1.0 dex, in agreement with other recent studies.
We derive a true distance modulus
to WLM of 25.144 $\pm$0.03 (r) $\pm$0.07 (s) mag from our data, in
good agreement with the earlier 24.92 $\pm$ 0.21 mag determination of 
Lee, Freedman and Madore (1993a) from Cepheid variables.  The quoted value
of the systematic uncertainty does not include the contribution from
the LMC distance which we have assumed to be 18.50 mag, as in the previous
papers in our project. 
\end{abstract}


\section{Introduction}
In our ongoing Araucaria Project, we are improving on the usefulness of a number
of stellar distance indicators by determining their environmental dependences from a study
of these objects in nearby galaxies with largely different environmental parameters.
We have described our approach and motivations
to improve the local calibration of the extragalactic
distance scale in a number of previous papers (Pietrzynski et al. 2002; Gieren et al. 2005a).
Among the known stellar methods of distance determination
Cepheid variables continue to be the most powerful standard candles to
determine the distances to galaxies out to about 10 Mpc, especially when they are used
in the near-infrared domain where the problems with dust absorption, particularly
intrinsic to the host galaxies, can be minimized (Gieren et al. 2005b, 2006; Pietrzynski
et al. 2006a; Soszynski et al. 2006). For this reason, we have made a considerable effort
to discover large samples of Cepheid variables in the target galaxies of the
Araucaria Project, viz. the irregular galaxies in the Local Group, and a number
of spiral galaxies in the nearby Sculptor Group (we are currently expanding our work
to several of the more massive spiral galaxies in both hemispheres, including M 83,
M 31 and M 81). Since Cepheid variables can be most easily
discovered in optical photometric bands where their light curves display the typical
sawtooth shapes and the amplitudes of the light variations are large, we have performed
extensive optical (VI) wide-field imaging surveys for Cepheid variables in all our target galaxies.
These surveys have discovered the first-ever reported Cepheids in the Sculptor galaxies
NGC 55 (Pietrzynski et al. 2006b), NGC 247 and NGC 7793 (in preparation), and have
greatly enhanced the number of known Cepheids with excellent light curves in the
optical V and I bands in the Local Group (NGC 6822: Pietrzynski et al. 2004; NGC 3109:
Pietrzynski et al. 2006c), and in the Sculptor Group spiral galaxy NGC 300 (Gieren et al. 2004).

The last of the irregular galaxies of the Local Group for which our project has not yet
provided a modern new survey for Cepheids is the WLM (Wolf-Lundmark-Melotte) galaxy.
In this paper, we report on the results of such an extensive survey which
has detected a large number of new Cepheid variables in WLM, greatly expanding the list of 15 Cepheid
variables which had been previously discovered by Sandage and Carlson (1985; hereafter SC85) 
from blue and yellow photographic plates
taken with the Palomar Hale 5-m and the Las Campanas 2.5-m du Pont reflectors between
1952 and 1983. All of the SC85 Cepheids in WLM have periods less than 10 days and their
usefulness for distance determination had been somewhat restricted therefore.
Our new Cepheid catalog given in section 3 of this paper contains 60 Cepheid
variables in WLM with periods down to 1.6 days, increasing the known Cepheid population
in WLM by a factor of four. Most importantly, we can also show from our
data that there is one truly long-period Cepheid (54 days) in WLM-the absence of 
such objects had been a major puzzle which was suspected to be a consequence of the low
metallicity of the young stellar population in WLM (SC85). For distance determination, 
the discovery of such a long-period Cepheid variable is clearly relevant and we will exploit 
on it in section 4 of this paper. 

Our paper is organized as follows. In section 2, we describe our observations, reductions
and calibrations. In section 3, we present the catalog of Cepheid variables discovered from
our wide-field images, including their periods and mean magnitudes. In section 4, we present
the period-luminosity (PL) relations resulting from our new data, and use these relations
to determine the distance to WLM. In section 5, we discuss our results, and in section 6 we summarize
the main conclusions of this paper.

\section{Observations,  Reductions and Calibrations}
All the  data presented in this paper were collected with the Warsaw 1.3-m 
telescope at Las Campanas Observatory. The telescope was equipped with 
a mosaic 8k $\times$ 8k detector, with a field of view of about 35 $\times$ 35 
arcmin and a scale of about 0.25 arcsec/pix. For more  instrumental
details on this camera, the reader is referred to the OGLE  website.\footnote{
{\it http://ogle.astrouw.edu.pl/\~{}ogle/index.html}}
V images of WLM were secured on 101 different nights between 
Dec 2001 and Dec 2003. In addition, 24 I-band images were collected 
in 2003. The exposure time was set to 900 seconds in both filters.

Preliminary reductions (i.e. debiasing and flatfielding)  were 
done with the IRAF\footnote{IRAF is distributed by the
National Optical Astronomy Observatories, which are operated by the
Association of Universities for Research in Astronomy, Inc., under cooperative
agreement with the NSF.} package. Then, PSF photometry was obtained 
for all stars in the same manner as described in Pietrzy{\'n}ski, Gieren and
Udalski (2002). Independently, the data were reduced with the OGLE III pipeline 
based on the image subtraction technique (Udalski 2003; Wo{\'z}niak 2000).  

In order to accurately calibrate our photometry onto the standard system 
our target was monitored during three photometric nights in 2000 - 2006,
each time together with a large number of photometric standard stars. 

On September 9, 2000 WLM was observed with the OGLE II instrumental system
(e.g. a one chip 2048 x 2048 pixel CCD camera) together with 
some 20 standards from the Landolt fields  spanning a wide range of
colors ( -0.14 $<$ V-I $<$ 1.54), and observed at very different
airmasses. The transformation from the OGLE II system onto the Landolt 
system was extremely well established from extensive observations 
of a large number of standard stars over several observing seasons
in the course of the OGLE II project  (e.g. Udalski et al. 1998, 2000).
Therefore, we adopted the following  transformations:\\

$$V-I= 0.969\times (v-i) + {\rm const}_{V-I}$$
$$V=v-0.002\times (V-I) + {\rm const}_V$$
$$I=i+0.029\times (V-I) + {\rm const}_I \eqno{(1)}$$

\noindent where the lower case letters {\it v,i} denote the aperture
instrumental magnitudes normalized to 1 sec exposure time, and the color coefficients 
are those derived by the OGLE team. The extinction coefficients and zero points
were derived from our data. The residuals did not exceed 0.04 mag (see Fig. 1) 
and did not show any significant dependence on color or magnitude. The accuracy of the 
zero points is estimated to be better than 0.02 mag in both filters.

In order to check the color coefficients provided by the OGLE team we 
also derived a full set of transformation coefficients. Then the instrumental 
magnitudes were transformed using our coefficients and the OGLE ones, 
and the results were compared. The resulting differences in magnitude 
were smaller than 0.007 mag in each band.

We also observed WLM together with a large set of Landolt standards 
covering a large range of colors ( -0.14 $<$ V-I $<$ 1.95) and at widely different air 
masses with the OGLE III mosaic camera on the same 1.3 m Warsaw telescope
on two different photometric nights.
WLM was located on two of the 8 chips of the camera (chip 2 and chip 3). 
Since in principle the transformation equations for each
chip may have different color coefficients and zero points, the selected
sample of standard stars was observed on each of the individual chips, and
transformation coefficients  were derived independently for each chip, 
on each night. 
The following transformations were obtained for chips 2 and 3:\\

$$V-I= 0.939\times (v-i) + {\rm const}_{V-I}$$
$$V=v-0.032\times (V-I) + {\rm const}_V$$
$$I=i+0.031\times (V-I) + {\rm const}_I ~~  chip 2 \eqno{(2)}$$

$$V-I= 0.936\times (v-i) + {\rm const}_{V-I}$$
$$V=v-0.030\times (V-I) + {\rm const}_V$$
$$I=i+0.037\times (V-I) + {\rm const}_I  ~~ chip 3 \eqno{(3)}$$

The resulting color coefficients are consistent with those derived
to calibrate our mosaic data from the same telescope and camera
for NGC 6822, NGC 3109 and NGC 55, other galaxies studied in 
the Araucaria Project (see the references given in the Introduction).
It is worth noticing that the 
color coefficients in equations 1, 2 and 3 are very small, showing
that both instrumental systems adjust very closely to the 
standard Cousins system.

To correct the possible small variation of the photometric zero points in V and I
over the mosaic, the "correction maps" established by Pietrzynski
et al. (2004) were used. These maps were already applied to correct our
photometry obtained in the field of NGC 6822 (Pietrzynski et al.
2004) and NGC 3109 (Pietrzynski et al. 2006c). Comparison with
other studies given in these papers revealed that these maps allow to correct
the zero point variations down to a level of better than 0.03 mag.

The differences between the mosaic camera zero points obtained on the two
independent photometric nights 
were found smaller than 0.03 mag, in  each filter and for both chips. 
In addition, the comparison of the photometry obtained with the OGLE II 
and OGLE III instrumental systems revealed that the 
differences in the zero points in both V and I bands  
are smaller than 0.02 mag and do not correlate in any significant way  
with magnitude or color (see Fig 1). As a result of all this comparative
work, we are sure that the  V and I magnitudes from the two cameras used in this study
are consistent at the 1-2\% level. This can also be seen in the quality
of the Cepheid light curves, particularly for the brightest variable, presented
in Fig. 2.

In order to perform an external check of our photometry we compared it 
to the recent results obtained by McConnachie et al. (2005), who 
kindly provided us with their data. 
While the zero point difference 
in the V band is reassuringly small (about 0.02 mag, within the errors), the mean difference 
in I amounts to 0.22 mag, in the sense that our I magnitudes are fainter
by this amount than the corresponding McConnachie et al. magnitudes for WLM stars. Also, there
is a clear color trend in the sense that for blue stars, our and McConnachie's I-band
magnitudes agree very well, but for redder stars there is an increasing discrepancy
with McConnachie's magnitudes becoming increasingly brighter than ours.
Since all the external checks on 
our I-band magnitudes obtained with the same telescope and cameras in our previously
studied Araucaria target galaxies always yielded good agreement with the photometry
of other authors, and since our mosaic camera I-band photometry agrees extremely well
with the OGLE II single-chip photometry which is calibrated to better than
1\%, we conclude that there must be a problem with the I-band data of
McConnachie et al. whose origin remains unknown to us, but could be related 
to the use of a non-standard I filter in their work.
Unfortunately, we are
not aware of any other source of I-band photometry for WLM we could 
directly compare our data with. Therefore we constructed the I band 
luminosity function for RGB stars in WLM using our new photometry, and 
measured the TRGB magnitude to be 20.91 $\pm$ 0.08 mag. This result is 
in good agreement with the I band TRGB magnitude determinations obtained 
for this galaxy by Lee et al. (1993b;  20.85 $\pm$ 0.05 mag), and 
by Minniti and Zijlstra (1997; 20.80 $\pm$ 0.05 mag), indicating 
that our present I band photometric zero point has been correctly determined,
within the stated small uncertainties. From the internal and external checks we
have made, any systematic zero point error on our present I-band magnitudes 
is limited to less than 2\%.

\section{Cepheid Catalog}

All stars identified in our photometry of WLM were searched for photometric variations with
periods between 0.2 and 100 days, using the analysis of variance algorithm 
(Schwarzenberg-Czerny 1989). In order to distinguish the Cepheids from other types
of variable stars we applied the criteria defined by Pietrzy{\'n}ski et al. (2002).
The light curves of all the variables identified as Cepheid candidates were fit by Fourier
series of order 2. We then rejected those objects with V amplitudes smaller than
0.4 mag, in agreement with the procedure we applied in the previous studies in this series.
In principle, one might expect a few very low-amplitude Cepheids close to the center of the
Hertzsprung progression (in the period range 10-13 days, depending on the metallicity),
or located close to the red edge of the Cepheid instability
strip. These latter Cepheids have decreased amplitudes due to the increased efficiency
in the convective energy transport (Bono et al. 2000) but seem to be normal with 
respect to their luminosities, and are therefore in principle useful for distance
determinations via PL relations. In our database for WLM we found only two such objects,
but we decided to omit them for the distance analysis because of the poor quality 
of their light curves, and in order to be consistent
with our earlier studies. In any case, including these objects would not change any
of the results and conclusions reached in this paper.

For the 60 stars passing our selection criteria, mean V and I magnitudes were derived
by integrating their light curves which had been previously converted onto an intensity
scale, and converting the results back onto the magnitude scale. The periods of the
60 Cepheids in our catalog range between 1.6 and 54.2 days. The accuracy of the period values 
is about $10^{-4}*P$ days. There is only one truly
long-period Cepheid in WLM, variable cep001 in our catalog, which is given in Table 1. The
variable with the next-longest period, cep002, has a period of 10.3 days. Since our
images cover the spatial content of WLM to almost 100\%, it seems that there are definitively
no other Cepheids with periods longer than 10.3 days in WLM except cep001. This bright
variable had already been discovered and classified as a Cepheid by SC85 (their variable
V12; entries in the "remarks" column in Table 1 give the Cepheid identifications of
SC85), but due to their limited set of photographic data they determined a wrong period
for this Cepheid (7.9 days). For all the other variables classified as Cepheids by
SC85, their Cepheid nature is confirmed in our study, although for most variables
the new periods differ quite significantly from the values given in SC85 which is not
a surprise given the quality and quantity of our new data, compared to the data SC85
had at their disposal for their very important and pioneering study of the stellar content
of WLM. We remark that from the 15 Cepheids discovered by SC85, two objects are not in
our catalog: V40 and V67. The variable V67 of SC85 falls between the chips of the
mosaic camera we used to image the WLM galaxy, and we have therefore no data for this object.
 Variable V40 shows a Cepheid-like
light curve from our data with a period close to the one found by SC85, but its amplitude
is below our threshold value which explains why it has not entered our catalog. The
low amplitude of this variable is also evident in its B light curve shown in SC85.

A comparison of the light curves of the 14 Cepheids in WLM common with SC85 shows a
dramatic increase in quality, and therefore in the accuracy of the periods and mean intensity
magnitudes we were able to derive from our new data. In Fig. 2, we show the V- and I light
curves for several of the Cepheids in our database whose quality is representative for
other Cepheids of similar periods. One can see that down to a period of 3 days, corresponding
to a mean V magnitude of about 22.2, the light curves are still very well defined and allow
the determination of the mean V magnitude with a precision of about 0.05 mag. The mean I magnitudes
of the Cepheids are somewhat less precise due to the smaller number of datapoints.

In Fig. 3, we show the locations of the WLM Cepheids in the V, V-I color-magnitude diagram,
where they delineate the expected Cepheid instability strip (Chiosi et al. 1992; Simon and Young 1997). 
The locations of the 60 variables in our catalog in the CMD lends further support to their correct
identifications as classical Cepheids. Again, it is remarkable to see the only long-period
variable in this diagram, at a V magnitude almost 2 mag brighter than all the other shorter-period,
lower-mass Cepheids we see in WLM. It still seems a challenging problem for stellar
evolution theory to explain the existence of just one high-mass star in the Cepheid instability
strip, given the rather abundant population of young, massive stars in this galaxy,
as evidenced by its blue supergiant population (Bresolin et al. 2006).

In Table 2, we report the individual V and I observations of the Cepheids in Table 1. The
full Table 2 is available in electronic form.

\section{PL relations and distance determination}

In Fig. 4, we show the V-band PL relation resulting from the data in Table 1, for the
Cepheids in our sample with logP (days) $>$ 0.5. The corresponding I-band PL relation defined
by these stars is shown in Fig. 5. These (35) objects represent the subsample
which should in principle be free of the Malmquist bias
which is introduced by retaining Cepheid variables close to the faint magnitude
cutoff of the photometry. This bias, if not accounted for, 
would tend to systematically decrease the derived distance to the galaxy. A sample of Cepheids
with logP (days) $>$ 0.5 should also be reasonably free of first overtone pulsators, whose existence 
at very short pulsation periods was impressively shown in the LMC work of the 
OGLE II Project (Udalski et al. 1999). 
Keeping only Cepheid variables above this period cutoff also assures, in the case
of the present photometry, that only Cepheid light curves of high quality are used for
the distance analysis.

A closer inspection of the sample of Cepheids in Figs. 4 and 5 reveals three objects
which are clearly too bright for their respective periods. These objects are
cep028 and cep031 which stand out in both PL planes, and the star cep032 which seems
too bright for its period in the I-band PL plane. These stars are likely to be strongly blended by
nearby companion stars which are not resolved in our photometry, but they could also be
overtone pulsators which at these periods near 4 days are still ocurring,
albeit in small numbers. In view
of this we decided to choose the cutoff period, in the case of WLM, at logP (days) = 0.7
(5.0 days). This choice yields the best compromise between retaining a statistically
significant sample of stars for the determination of the PL relations (19 Cepheids), and avoiding
a possibly significant contamination of the sample by overtone Cepheids and and/or heavily
blended objects. As we will show below, the distance determination to WLM is however not
significantly affected by the choice of the cutoff period-both samples, using logP = 0.5, or 0.7
as the cutoff period, yield distance moduli to WLM which differ by only 0.05 mag, which
is within the uncertainty of the present distance determination of WLM from our
Cepheid photometry.

>From an inspection of Figs. 4 and 5, it is obvious that the slopes of the PL relations
in V and I adopted from the LMC Cepheids as given by the OGLE II Project provide excellent
fits to the present data for the WLM Cepheids. Indeed, fits to a straight line to our data  
yield the following slopes for the PL relations: -2.57 $\pm$ 0.16, -2.93 $\pm$ 0.12 
and -3.15 $\pm$ 0.16 in V, I and ${\rm W_{\rm I}}$, the reddening-free Wesenheit band (see Fig. 6),
respectively. These values are consistent with the corresponding OGLE slopes of
the LMC Cepheid PL relation of -2.775, -2.977 
and -3.300 (Udalski 2000) at the level of 1 $\sigma$. In the case of WLM, the low number of long-period
Cepheids and the large gap in period between 10 and 54 days make an accurate determination
of the slope of the PL relations impossible, but the data are clearly very well fit by the slopes
adopted from the LMC Cepheids, supporting the conclusion that any systematic change of the PL
relation slope as going from the LMC Cepheids to the more metal-poor WLM Cepheids must be
very small. Our present data are certainly fully consistent with the assumption of identical slopes
of the Cepheid PL relations in V, I and ${\rm W_{\rm I}}$ for WLM and the LMC.

In view of this finding, we are justified to adopt the extremely well determined OGLE slopes to derive the 
distance to WLM, as we have already done in the previous papers of this series.
 This leads to the following equations: (logP (days) $>$ 0.7; 19 Cepheids):  \\

V = -2.775 log P + (23.772 $\pm$ 0.037) \\

I = -2.977 log P + (23.275 $\pm$ 0.028) \\

${\rm W}_{\rm I}$ = -3.300 log P + (22.418  $\pm$ 0.045) \\

Using the 35 Cepheids with logP (days) $>$ 0.5, we obtain the following results: \\

V = -2.775 log P + (23.722 $\pm$ 0.037) \\
                                                                                                  
I = -2.977 log P + (23.214 $\pm$ 0.030) \\

${\rm W}_{\rm I}$ = -3.300 log P + (22.382  $\pm$ 0.036) \\

Adopting, as in our previous papers, a value of 18.50 mag for the true distance modulus to the LMC, 
a value of E(B-V) = 0.02 mag for the foreground 
reddening toward WLM (Schlegel  et al. 1998), and the reddening law of 
Schlegel et al. (1998) ( ${\rm A}_{\rm V}$ = 3.24 E(B-V), ${\rm A}_{\rm I}$ = 1.96
E(B-V)) we obtain the following reddening-corrected distance moduli for WLM in the
three different bands:\\

1. (19 Cepheids with logP (days) $>$ 0.7):\\

$(m-M)_{0}$ (${\rm W}_{\rm I}$) = 25.144 mag \\

$(m-M)_{0}$ (I) = 25.142 mag\\

$(m-M)_{0}$ (V) = 25.050 mag\\

2. (35 Cepheids with logP (days) $>$ 0.5):\\

$(m-M)_{0}$ (${\rm W}_{\rm I}$) = 25.093 mag \\

$(m-M)_{0}$ (I) = 25.082 mag\\

$(m-M)_{0}$ (V) = 25.014 mag\\

As we already mentioned above, the difference in the true distance moduli in the respective bands,
for the two samples,
is in the order of the uncertainty on the zero points of the respective PL relation,
demonstrating that the choice of the period cutoff for the WLM Cepheid sample used
for the distance determination is not a source of significant systematic error on
the WLM distance. We adopt 25.144 $\pm$ 0.040 (random error) mag 
 as our best determination of the true distance
modulus of WLM from the reddening-independent V-I Wesenheit magnitudes of its Cepheids.
We will discuss this result, and estimate its total uncertainty in the following section.

\section{Discussion}
The current distance result for the WLM dwarf irregular galaxy is based on a sizeable sample of Cepheid
variables with excellent light curves
which have been mostly discovered in our present wide-field imaging survey. Very importantly,
we have discovered one long-period variable which allows a check on the slope of the Cepheid PL relation
in WLM and partly resolves the mystery of the absence of such stars in WLM previously discussed 
by Sandage and Carlson (1985). Our data in the period-mean magnitude planes in V, I and ${\rm W_{\rm I}}$
are very well fit
with the PL relation slopes obtained for the LMC Cepheids by the OGLE II project, further
supporting the evidence that the slope of the Cepheid PL relation is independent of metallicity
down to very low values of [Fe/H] or [O/H]. For WLM, the mean oxygen abundance of 3 blue
supergiant stars was recently measured to be about -0.6 dex (Bresolin et al. 2006), suggesting
that its older population of Cepheid variables is likely to have a mean metallicity close to -1.0 dex,
which is indeed considerably more metal-poor than the mean [Fe/H] of -0.34 dex 
derived for a sample of LMC Cepheids by Luck et al. (1998). This is in agreement with the result of Udalski
et al. (2001) for the Cepheid PL relation in another metal-poor Local Group dwarf irregular galaxy,
IC 1613, which also does not show any sign for a change of the slope of the PL relation
at very low metallicities. The recent results of Gieren et al. (2005c) from a comparison
of the Cepheid PL relations in the LMC and Milky Way, and of Macri et al. (2006) from a
comparison of the Cepheid PL relations for two fields of very different mean metal abundance in the maser
galaxy NGC 4258 observed with HST/ACS also support the constancy of the slope of the
Cepheid PL relation in optical bands up to solar metallicity. Very importantly, a completely
independent confirmation of this has very recently come from the HST parallaxes of a
number of nearby Milky Way Cepheids derived by Benedict et al. (2006) which also suggest
that there is no difference between the slope of the PL relation in the Milky Way galaxy
and the LMC.

The current distance determination to WLM is subject to the several well-known sources of systematic
uncertainty in such studies. We have made a great effort to calibrate our data as accurately
as possible, and our discussion in section 2 in this paper shows that we can confidently assume
that our photometric zero points in V and I are accurate to better than $\pm$0.03 mag. The
sample of Cepheid variables in our study is large enough to ensure that our distance result 
is not severely affected
by the problem of a possible incomplete filling of the instability strip. We do, however, note
that the range of periods for which a complete filling of the instability strip can be assumed
is rather limited (5-11 days). We recall that we are not attempting to use our data to fit
slopes to the PL diagrams in Figs. 4-6 whose values would sensitively depend on the exact position
of the one long-period Cepheid with respect to the ridge line in the instability strip,
and rather adopt the slopes from the LMC PL relations. 
It is reassuring to see that the long-period Cepheid in our sample falls very close
to the fitting lines in all filters, suggesting that this star is located close to the center
of the Cepheid instability strip. This conclusion is also clearly supported by the position 
of object cep001 in the color-magnitude diagram in Fig. 3. All the physical
information on this star available to us (shape of the light curve, amplitude, mean magnitude,
color, period) supports that this object is a normal classical Cepheid. Although it may appear
surprising, our database clearly indicates that there are no other Cepheids in the large period gap
between the objects cep001 and cep002 in WLM.

Our discussion in
the previous section has also shown that our adopted choice of the cutoff period, necessary
to exclude overtone pulsators from the sample and to address the problem of Malmquist bias,
is not affecting our distance result by more than $\pm$0.05 mag. The small dispersion of
the data points in 
Fig. 6 around the mean PL relation suggests that 
the process of eliminating the influence of reddening (both foreground,
and a possible additional variable reddening produced inside WLM itself)
by the construction of the reddening-free Wesenheit magnitudes of the Cepheids has worked very well.
This is likely due to the fact
that our Cepheid photometry is very little affected by blending with unresolved, relatively bright nearby
companion stars which is a more serious problem in the spiral galaxies of higher stellar density
in our program. Yet, even in the case of NGC 300, at about twice the distance of WLM (Gieren et al. 2005b),
Bresolin et al. (2005) were able to show from a comparison of ground-based data to HST/ACS data
that the effect of blending on the distance modulus is less than 0.04 mag. We therefore believe
that the distance modulus of WLM from the Wesenheit PL relation eliminates reddening as a
significant source of systematic error in our study. Also, the fact that the distance result
from the I band is practically identical to the one from the Wesenheit band seems to indicate
that the total reddening affecting the WLM Cepheids in our database is very small, which
in turn means that in addition to the very small 0.02 mag foreground reddening there is very little
additional dust absorption {\it intrinsic} to WLM. A follow-up study of the WLM Cepheids in the near-IR
J and K bands will shed more light on this and allow an accurate determination of any residual
dust absorption inside WLM, as we have done in our previous studies in several of the target
galaxies of our project. Unfortunately, so far we have not been able to collect such near-IR images
of WLM under photometric conditions but hope to do so in the near future.

As a conclusion, our present distance modulus determination of WLM from the Wesenheit magnitudes
of its long(er)-period Cepheid population has a total estimated systematic uncertainty
of $\pm$0.07 mag, when the different contributions discussed about are added in quadrature.
Therefore, we obtain as our best result from the current study a true distance modulus of 
the WLM galaxy of 25.144 $\pm$0.03 (random) $\pm$0.07 (systematic) mag, equivalent to a total
uncertainty of $\pm$4\%. This estimation of the total uncertainty does {\it not} include,
however, the uncertainty on our adopted value of 18.50 mag for the distance of the LMC. A thorough
discussion of this value will be provided in a forthcoming paper once the Cepheid distances
to all Araucaria project target galaxies have been measured. At this point, we just mention
that the recent absolute calibration of the Cepheid PL relation of Macri et al. (2006) in NGC 4258 which
is tied to the geometrical maser distance to this galaxy implies a LMC distance modulus
of 18.41, but the uncertainty on this value estimated by the authors of that paper
 makes it clearly compatible with our adopted
LMC distance of 18.50. It seems clear that the adopted distance to the LMC continues to be
the largest individual source of systematic error on modern Cepheid-based distance determinations
to nearby galaxies
like the present one, which have succeeded in beating down other systematics to a few percent level.
Work to improve this situation will be extremely important over the next years. 
As a positive note, evidence is now clearly mounting that the {\it slope}
of the PL relation is independent of metallicity over the broad range from solar down to
about -1.0 dex allowing us to use the slope values determined in the LMC
by the microlensing projects with confidence for Cepheid-based distance determinations to other galaxies,
including those showing pronounced radial metal abundance variations in their disks.

Finally, we note that our improved Cepheid  distance determination to WLM puts the galaxy
some 0.2 - 0.3 mag
further away than the value derived from the TRGB I band magnitude
(Lee, Freedman and Madore (1993b): - 24.87 $\pm$ 0.08 ; Minniti and Zijlstra (1997): - 24.75
$\pm$ 0.1 ; this paper: - 24.91 $\pm$ 0.08). A possible interpretation of this discrepancy
is that the metal-poor WLM Cepheids are, at a given period, intrinsically fainter in V and I than
their more metal-rich counterparts in the LMC. Indeed, the sign
and size of the discrepancy between the present distance to WLM from its Cepheids, and the one derived
from the TRGB magnitude is consistent with the metallicity dependence of the zero point
of the Cepheid PL relation of 0.2-0.3 mag/dex found by Sakai et al. (2004-their Fig. 15).
On the other hand, our current work on the distances of nearby galaxies from the blue
supergiant Flux-Weighted Gravity-Luminosity Relation (FGLR; Kudritzki, Bresolin and Przybilla 2003)
supports the 25.14 mag true distance modulus derived in this paper from the WLM Cepheids
(Kudritzki et al. 2007, in preparation), so the interpretation that the Cepheid distance
to WLM is longer than the TRGB distance because of a metallicity effect on the PL
relation zero point may be premature at this time.
A full discussion of the effect of metallicity on the zero point of the Cepheid PL relation,
in different bands, will be presented in a later stage of our project when distances
from a variety of methods to all target galaxies of the project will have been determined,
and will hopefully lead to a very accurate calibration of the metallicity dependence
of the PL relation in various photometric bands.

\section{Conclusions}
The main conclusions of this paper can be summarized as follows:

1. We have conducted an extensive wide-field imaging survey for Cepheids in the Local Group
dwarf irrgular galaxy WLM. From V-band images obtained on 101 different nights, we have
found 60 Cepheids with periods down to 1.6 days. Down to a period of
about 3 days, our Cepheid survey in WLM
should be essentially complete. We have determined accurate periods and mean magnitudes for all
variables in the V, I and Wesenheit bands.

2. We have discovered the first (and only) long-period Cepheid variable in WLM, cep001 in our
catalog, with a period of 54.2 days. This variable had already been discovered before by Sandage and Carlson (1985),
but their low-quality data had led them to derive a wrong period for this Cepheid.

3. From the data in our catalog we have constructed PL relations in the V, I and the reddening-independent
Wesenheit band. We find that our data are very well fit by the slopes of the corresponding 
PL relations determined in the LMC by the OGLE II project, supporting the conclusion that
the slope of the PL relation defined by the more metal-poor Cepheids in WLM is identical
to the one in the LMC.

4. We have derived absorption-corrected distance moduli to WLM from the data in V, I and W. Our adopted best value
for the WLM distance modulus from the reddening-independent Wesenheit magnitudes of the Cepheids
is 25.144 $\pm$0.03 (random) $\pm$0.07 (systematic) mag. The excellent agreement between the
W-band and I-band distance modulus values hints at very little dust absorption intrinsic to WLM.

5. The total uncertainty of our present distance determination of $\sim$$\pm$4\% does not include
the contribution from the uncertainty on the adopted LMC distance of 18.50 to which
our present distance determination to WLM is tied, as in the previous papers of the
Araucaria project. As in our previous Cepheid studies of NGC 6822
(Gieren et al. 2006), IC 1613 (Pietrzynski et al. 2006a), NGC 3109 (Pietrzynski et al. 2006c, 
Soszynski et al. 2006), NGC 300 (Gieren et al. 2004; Gieren et al. 2005b), and NGC 55 (Pietrzynski
et al. 2006b), the total error on our Cepheid distance to WLM due to the variety of factors discussed in
the previous section
is clearly smaller than the contribution coming from the adopted LMC distance, implying that the
main obstacle to significant progress in the measurement 
of the {\it absolute} distances to nearby galaxies is our continuing difficulty to obtain a
truly high-quality measurement of the distance to the LMC. 

6. With WLM, there is now another galaxy in our project whose Cepheid distance can be compared
to the distances we will measure for our target galaxies from a variety of other methods, like the
Flux-Weighted Gravity-Luminosity relationship of Kudritzki et al. (2003) for blue
supergiant stars. It is another step toward the main goal of the Araucaria Project, 
viz. an accurate
determination of the environmental dependences of different stellar distance indicators, with
the corresponding reduction on the systematic error on the Hubble constant determined from
secondary distance indicators which will be re-calibrated from the standard candles we are
investigating once their environmental dependences are well established.

\acknowledgements
We are grateful to the staff of Las Campanas 
Observatory, and to the CNTAC for providing the large amounts of 
telescope time which were necessary to complete this project.
GP, WG, DM, RM and AG gratefully acknowledge 
financial support for this
work from the Chilean Center for Astrophysics FONDAP 15010003.    
Support from the Polish grant N203 002 31/0463 is also acknowledged.

\begin{deluxetable}{c c c c c c c c c}
\tablecaption{Cepheids in WLM}
%\tablewidth{0pt}
\tablehead{
\colhead{ID} & \colhead{RA} & \colhead{DEC}  & \colhead{P} & \colhead{ ${\rm
T}_{0}$} &
\colhead{$<V>$} & \colhead{$<I>$} & \colhead{$<W_{\rm I}>$} & \colhead{remarks}\\
 & \colhead{(J2000)} & \colhead{(J2000)}  &
\colhead{ [days]} &  &
\colhead{[mag]} & \colhead{[mag]} & \colhead{[mag]} & 
}
\startdata
cep001 & 0:01:57.48 & -15:24:50.9 &  54.17118 & 2452200.59519 &  19.124 &  18.145 &  16.628 & V12\\
cep002 & 0:01:54.33 & -15:30:00.1 &  10.34249 & 2452201.65246 &  21.110 &  20.352 &  19.177 & \\
cep003 & 0:01:54.03 & -15:27:05.2 &  10.33645 & 2452202.34982 &  20.975 &  20.247 &  19.119 & \\
cep004 & 0:02:03.30 & -15:26:23.4 &  10.32152 & 2452200.25769 &  21.270 &  20.416 &  19.092 & \\
cep005 & 0:01:54.68 & -15:29:55.0 &   8.63110 & 2452194.74995 &  21.182 &  20.471 &  19.369 & V21\\
cep006 & 0:01:57.08 & -15:30:56.6 &   8.12579 & 2452195.29021 &  21.350 &  20.600 &  19.438 & V24\\
cep007 & 0:02:00.19 & -15:24:11.4 &   8.12051 & 2452200.62453 &  21.319 &  20.629 &  19.559 & \\
cep008 & 0:01:56.56 & -15:27:15.6 &   7.49672 & 2452202.09693 &  21.170 &  20.511 &  19.490 & V7\\
cep009 & 0:01:57.07 & -15:27:25.7 &   7.34322 & 2452199.48513 &  21.297 &  20.573 &  19.451 & V48\\
cep010 & 0:01:56.40 & -15:24:33.0 &   7.32485 & 2452198.47873 &  21.314 &  20.661 &  19.649 & V11\\
cep011 & 0:01:57.04 & -15:29:36.9 &   6.64055 & 2452201.99522 &  21.258 &  20.640 &  19.682 & V8\\
cep012 & 0:02:00.51 & -15:25:23.2 &   6.15754 & 2452201.17636 &  21.423 &  20.736 &  19.671 & \\
cep013 & 0:01:53.27 & -15:29:40.5 &   6.05309 & 2452197.85970 &  21.775 &  21.086 &  20.018 & V37\\
cep014 & 0:02:10.26 & -15:33:31.9 &   5.56746 & 2452197.36270 &  21.557 &  20.973 &  20.068 & \\
cep015 & 0:01:59.91 & -15:24:49.4 &   5.43153 & 2452200.95046 &  21.544 &  21.043 &  20.266 & V50\\
cep016 & 0:01:55.20 & -15:24:26.8 &   5.20796 & 2452201.89881 &  21.760 &  21.090 &  20.052 & \\
cep017 & 0:01:57.32 & -15:29:03.9 &   5.12851 & 2452201.58810 &  21.963 &  21.281 &  20.224 & \\
cep018 & 0:01:56.16 & -15:25:44.8 &   5.02134 & 2452197.96031 &  21.964 &  21.211 &  20.044 & \\
cep019 & 0:02:00.12 & -15:25:15.1 &   4.92341 & 2452199.88515 &  21.507 &  21.009 &  20.237 & \\
cep020 & 0:01:59.50 & -15:25:57.4 &   4.91559 & 2452201.10843 &  21.708 &  21.002 &  19.908 & V29\\
cep021 & 0:02:01.47 & -15:23:20.1 &   4.86831 & 2452199.59021 &  21.993 &  21.356 &  20.369 & \\
cep022 & 0:02:00.20 & -15:25:17.2 &   4.71140 & 2452198.82984 &  22.012 &  21.278 &  20.140 & \\
cep023 & 0:02:00.85 & -15:25:04.9 &   4.61967 & 2452197.94467 &  22.181 &  21.371 &  20.116 & \\
cep024 & 0:01:52.13 & -15:27:05.8 &   4.36958 & 2452198.61530 &  21.904 &  21.318 &  20.410 & V66\\
cep025 & 0:02:02.37 & -15:23:36.9 &   4.05416 & 2452199.04623 &  22.231 &  21.597 &  20.614 & \\
cep026 & 0:02:02.23 & -15:26:45.1 &   3.97606 & 2452199.52316 &  22.375 &  21.490 &  20.118 & \\
cep027 & 0:01:55.96 & -15:26:22.1 &   3.86503 & 2452199.27326 &  22.262 &  21.546 &  20.436 & \\
cep028 & 0:01:55.92 & -15:29:02.9 &   3.83017 & 2452201.20424 &  21.528 &  20.990 &  20.156 & \\
cep029 & 0:02:01.15 & -15:32:09.2 &   3.82512 & 2452198.77320 &  22.076 &  21.478 &  20.551 & V38\\
cep030 & 0:01:51.94 & -15:27:11.1 &   3.74707 & 2452198.98222 &  22.129 &  21.479 &  20.472 & \\
cep031 & 0:01:54.57 & -15:25:18.1 &   3.65377 & 2452200.36249 &  21.606 &  21.044 &  20.173 & \\
cep032 & 0:02:00.60 & -15:26:21.8 &   3.47735 & 2452201.41880 &  22.033 &  21.248 &  20.031 & \\
cep033 & 0:01:58.99 & -15:24:37.3 &   3.30475 & 2452199.14958 &  22.193 &  21.655 &  20.821 & \\
\enddata
\end{deluxetable}

\setcounter{table}{0}
\begin{deluxetable}{c c c c c c c c c}
\tablecaption{Cepheids in WLM - continued}
%\tablewidth{0pt}
\tablehead{
\colhead{ID} & \colhead{RA} & \colhead{DEC}  & \colhead{P} & \colhead{
${\rm T}_{0}$} &
\colhead{$<V>$} & \colhead{$<I>$} & \colhead{$<W_{\rm I}>$} & \colhead{remarks}\\
 & \colhead{(J2000)} & \colhead{(J2000)}  &
\colhead{ [days]} &  &
\colhead{[mag]} & \colhead{[mag]} & \colhead{[mag]} &
}
\startdata
cep034 & 0:02:00.46 & -15:26:44.2 &   3.24559 & 2452199.75819 &  22.107 &  21.536 &  20.651 & \\
cep035 & 0:01:58.10 & -15:28:57.9 &   3.20292 & 2452199.69890 &  22.106 &  21.620 &  20.867 & \\
cep036 & 0:01:59.18 & -15:24:18.1 &   3.13908 & 2452200.37271 &  22.075 &  21.700 &  21.119 & V1\\
cep037 & 0:02:00.13 & -15:24:20.8 &   3.13102 & 2452199.64936 &  22.200 &  21.582 &  20.624 & \\
cep038 & 0:02:01.47 & -15:25:08.9 &   3.03891 & 2452199.44801 &  22.301 &  21.544 &  20.371 & \\
cep039 & 0:01:58.00 & -15:23:39.5 &   3.02576 & 2452201.61526 &  22.413 &  21.669 &  20.516 & \\
cep040 & 0:02:04.72 & -15:25:11.0 &   2.96301 & 2452199.33434 &  21.907 &  21.272 &  20.288 & \\
cep041 & 0:01:50.67 & -15:28:55.4 &   2.92899 & 2452202.05423 &  22.550 &  21.888 &  20.862 & \\
cep042 & 0:02:01.68 & -15:32:13.8 &   2.92454 & 2452202.46634 &  22.289 &  21.804 &  21.052 & \\
cep043 & 0:01:58.03 & -15:30:38.6 &   2.88896 & 2452201.16317 &  22.002 &  21.462 &  20.625 & \\
cep044 & 0:01:57.18 & -15:31:30.1 &   2.83459 & 2452199.52500 &  22.028 &  21.468 &  20.600 & \\
cep045 & 0:01:54.73 & -15:22:29.8 &   2.80077 & 2452201.45090 &  21.926 &  21.397 &  20.577 & \\
cep046 & 0:01:57.84 & -15:22:09.3 &   2.74179 & 2452200.78817 &  22.976 &  22.500 &  21.762 & \\
cep047 & 0:02:01.59 & -15:19:50.0 &   2.64641 & 2452200.62767 &  22.971 &  22.208 &  21.025 & \\
cep048 & 0:02:09.54 & -15:22:53.6 &   2.52010 & 2452201.90797 &  22.474 &    - &   -  & \\
cep049 & 0:01:59.67 & -15:25:16.3 &   2.51004 & 2452200.24177 &  22.055 &  21.729 &  21.224 & \\
cep050 & 0:02:00.74 & -15:28:58.4 &   2.48031 & 2452201.87459 &  21.575 &  20.958 &  20.002 & \\
cep051 & 0:01:58.61 & -15:26:11.6 &   2.38903 & 2452199.88267 &  21.533 &  20.631 &  19.233 & \\
cep052 & 0:02:03.77 & -15:28:49.7 &   2.38701 & 2452200.11112 &  22.001 &  21.567 &  20.894 & \\
cep053 & 0:02:03.63 & -15:32:42.7 &   2.35613 & 2452201.71542 &  22.916 &  22.540 &  21.957 & \\
cep054 & 0:02:03.22 & -15:24:55.6 &   2.34579 & 2452200.82716 &  22.761 &  22.053 &  20.956 & \\
cep055 & 0:02:00.77 & -15:26:25.5 &   2.33770 & 2452200.24656 &  21.459 &  21.118 &  20.589 & \\
cep056 & 0:02:09.55 & -15:31:35.0 &   2.29048 & 2452200.50653 &  22.908 &  22.298 &  21.352 & \\
cep057 & 0:01:59.85 & -15:28:59.8 &   2.16776 & 2452201.55259 &  22.877 &  22.046 &  20.758 & \\
cep058 & 0:01:54.61 & -15:28:37.0 &   2.12863 & 2452201.26721 &  22.358 &  21.748 &  20.802 & \\
cep059 & 0:02:00.00 & -15:25:59.8 &   1.64746 & 2452201.88190 &  22.395 &  21.315 &  19.641 & blend\\
cep060 & 0:01:53.57 & -15:29:57.0 &   1.62627 & 2452201.78043 &  22.640 &  22.234 &  21.605 & \\
\enddata
\end{deluxetable}

\begin{deluxetable}{ccccc}
\tablecaption{Individual V and I Observations}
\tablehead{
\colhead{object}  & \colhead{filter} &
\colhead{HJD-2450000}  & \colhead{mag}  & \colhead{$\sigma_{mag}$}\\
}
\startdata
cep001 & V & 2859.767260 &  18.992 &   0.011\\
cep001 & V & 2870.773910 &  19.139 &   0.010\\
cep001 & V & 2877.719380 &  19.311 &   0.011\\
cep001 & V & 2884.726910 &  19.531 &   0.011\\
cep001 & V & 2902.671760 &  18.838 &   0.007\\
cep001 & V & 2906.700510 &  18.876 &   0.008\\
cep001 & V & 2910.673090 &  18.946 &   0.010\\
cep001 & V & 2915.629040 &  19.010 &   0.012\\
cep001 & V & 2930.643450 &  19.237 &   0.011\\
cep001 & V & 2934.624860 &  19.378 &   0.011\\
cep001 & V & 2942.576190 &  19.560 &   0.014\\
cep001 & V & 2950.543020 &  18.961 &   0.019\\
cep001 & V & 2954.584470 &  18.837 &   0.014\\
cep001 & V & 2959.548090 &  18.884 &   0.007\\
cep001 & V & 2963.554610 &  18.947 &   0.010\\
cep001 & V & 2966.523340 &  18.979 &   0.010\\
cep001 & V & 2968.541650 &  19.018 &   0.011\\
cep001 & V & 2971.552980 &  19.051 &   0.010\\
cep001 & V & 2972.535090 &  19.053 &   0.011\\
cep001 & V & 2973.555780 &  19.058 &   0.012\\
\enddata
\tablecomments{The complete version of this table is in the electronic
edition of the Journal.  The printed edition contains only
the the first 20 measurements in V band for the Cepheid variable cep001.}

\end{deluxetable}

\begin{figure}[p]
\vspace*{25cm}
\includegraphics{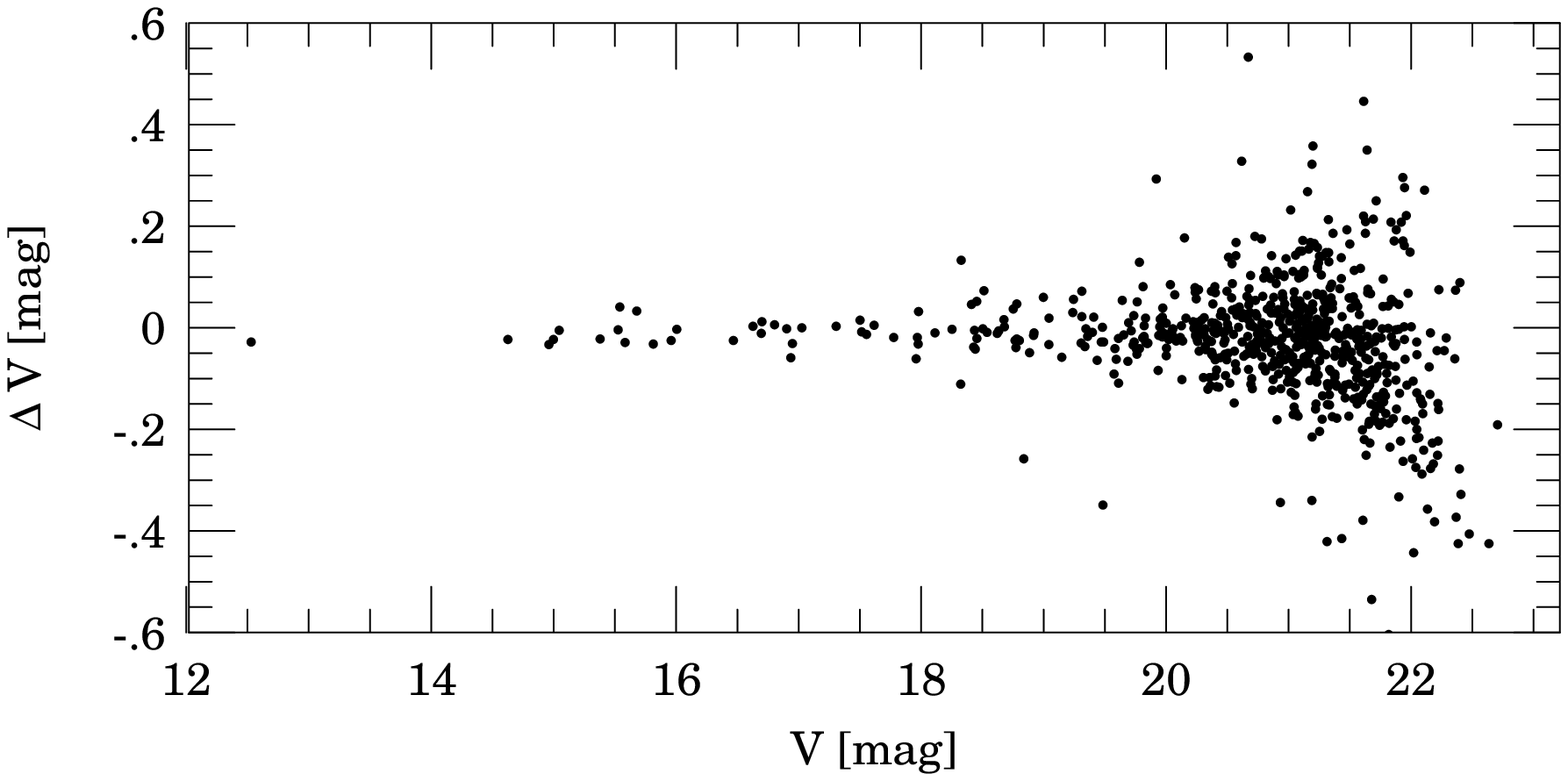}
\includegraphics{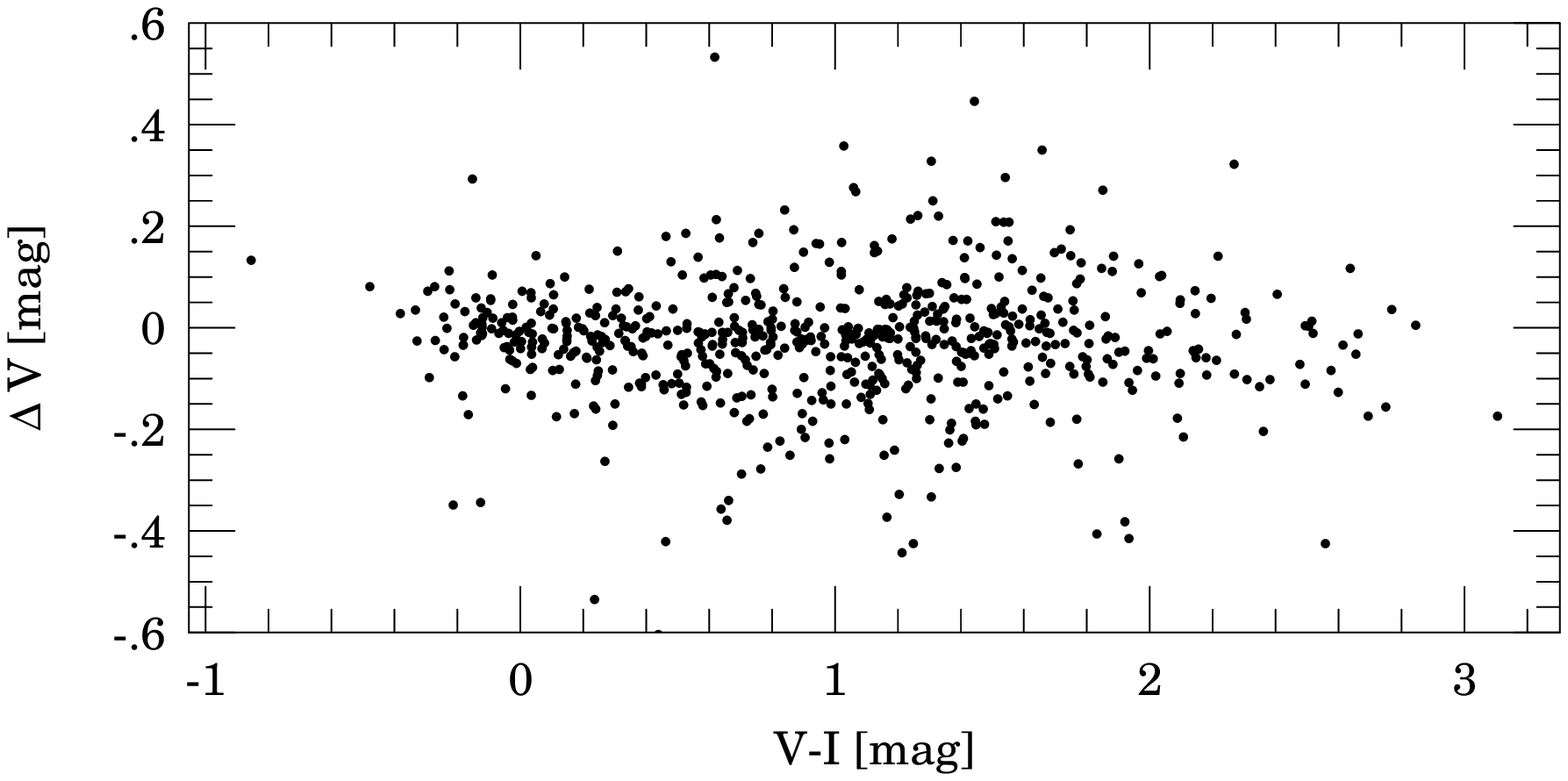}
\includegraphics{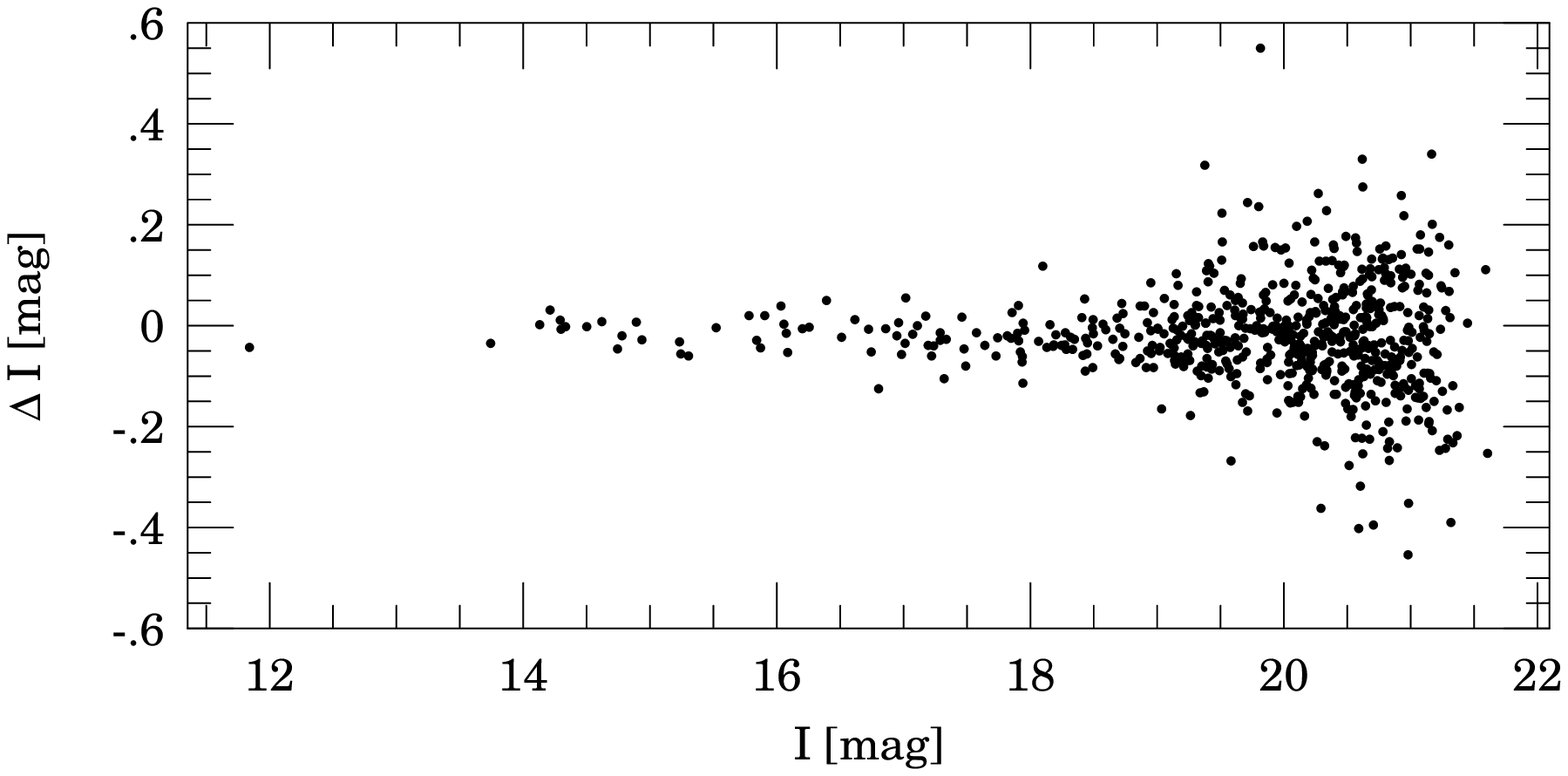}
\includegraphics{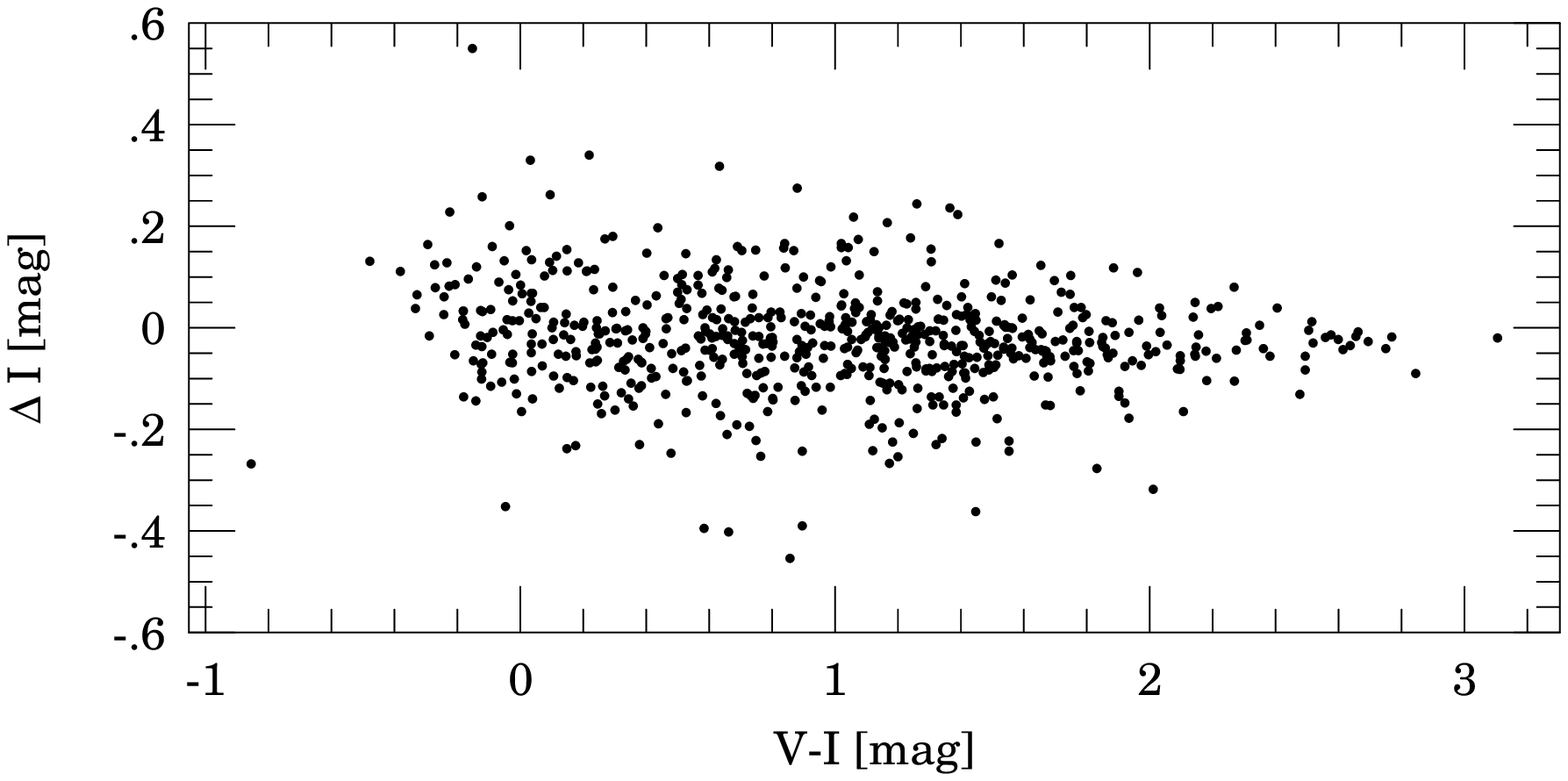}
\caption{The comparison of the photometric zero points in our V- and I-band
photometry of WLM obtained from the OGLE II single chip and the OGLE III
mosaic camera. The independent zero point determinations are consistent
to better than 0.02 mag, and there are no significant trends with color
or magnitude in the data.}
\end{figure}

\begin{figure}[htb]
\vspace*{22cm}
\includegraphics{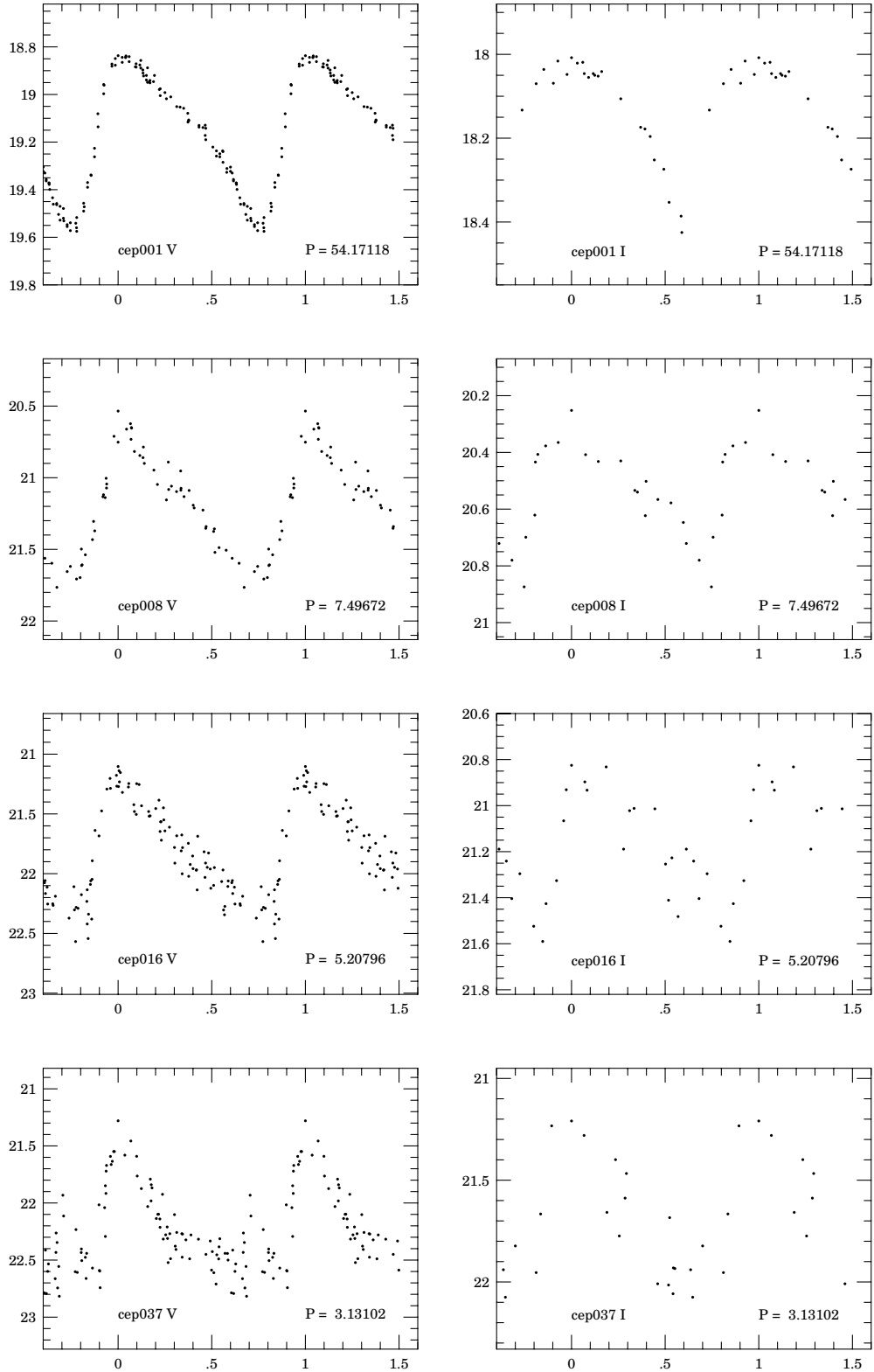}
\caption{Phased V- and I-band light curves for some Cepheids of different
periods in our WLM catalog. These light curves are representative for the
light curves of other Cepheid variables of similar periods.
}
\end{figure}

\begin{figure}[htb]
\vspace*{15cm}
\includegraphics{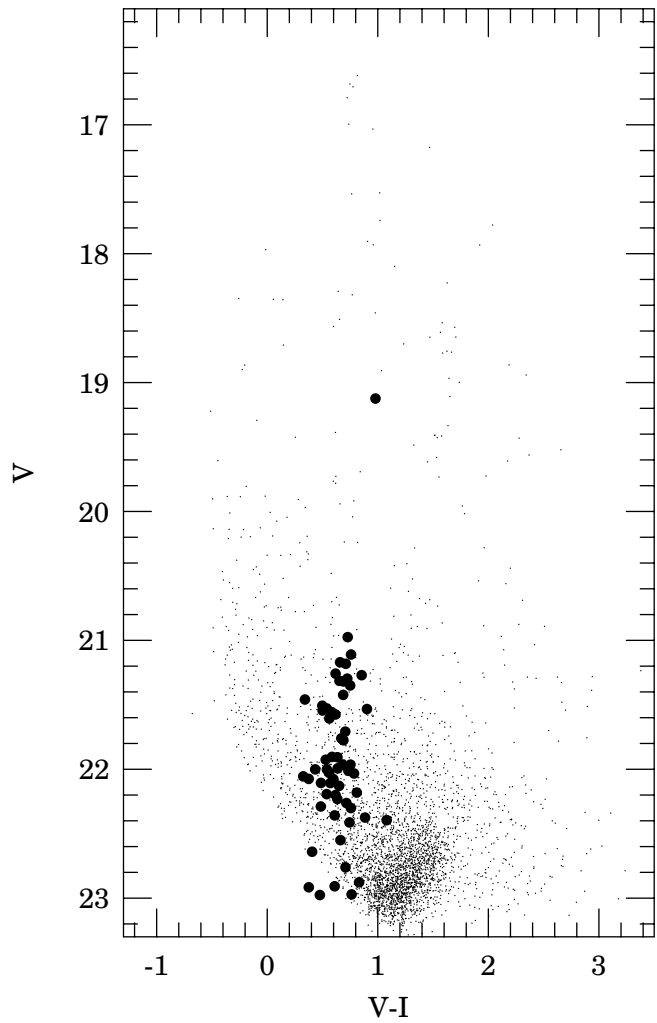}
\caption{The V,V-I magnitude-color diagram for WLM. The Cepheids discovered
in our survey are marked with black circles. They fill the expected
region of the Cepheid instability strip for fundamental mode pulsators 
in this diagram, yielding supporting evidence that the classification of
the variables as classical fundamental mode Cepheids is correct.
}
\end{figure}

\begin{figure}[htb]
\vspace*{15cm}
\includegraphics{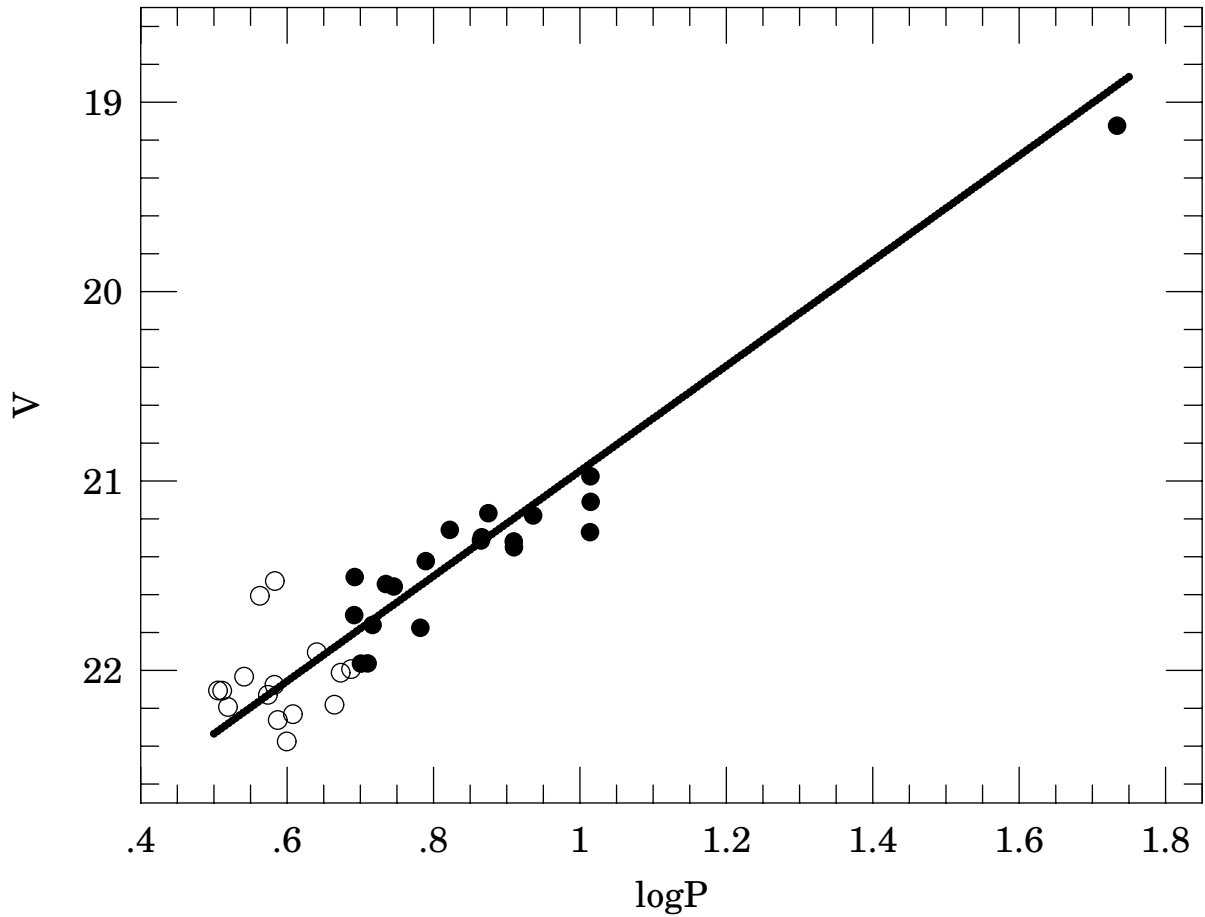}
\caption{The period-luminosity relation from the 35 Cepheid variables in WLM with 
logP (days) $>$ 0.5. Black circles show those Cepheids with logP $>$ 0.7.
This sample of 19 stars is unaffected by Malmquist bias and contamination with 
possible overtone pulsators, and the mean magnitudes of the variables are determined to better
than 1\% (random uncertainty). The slope of the fitting line is taken from
the LMC Cepheid PL relation of the OGLE II project and provides an excellent fit
to the data.
}
\end{figure}

\begin{figure}[htb]
\vspace*{15cm}
\includegraphics{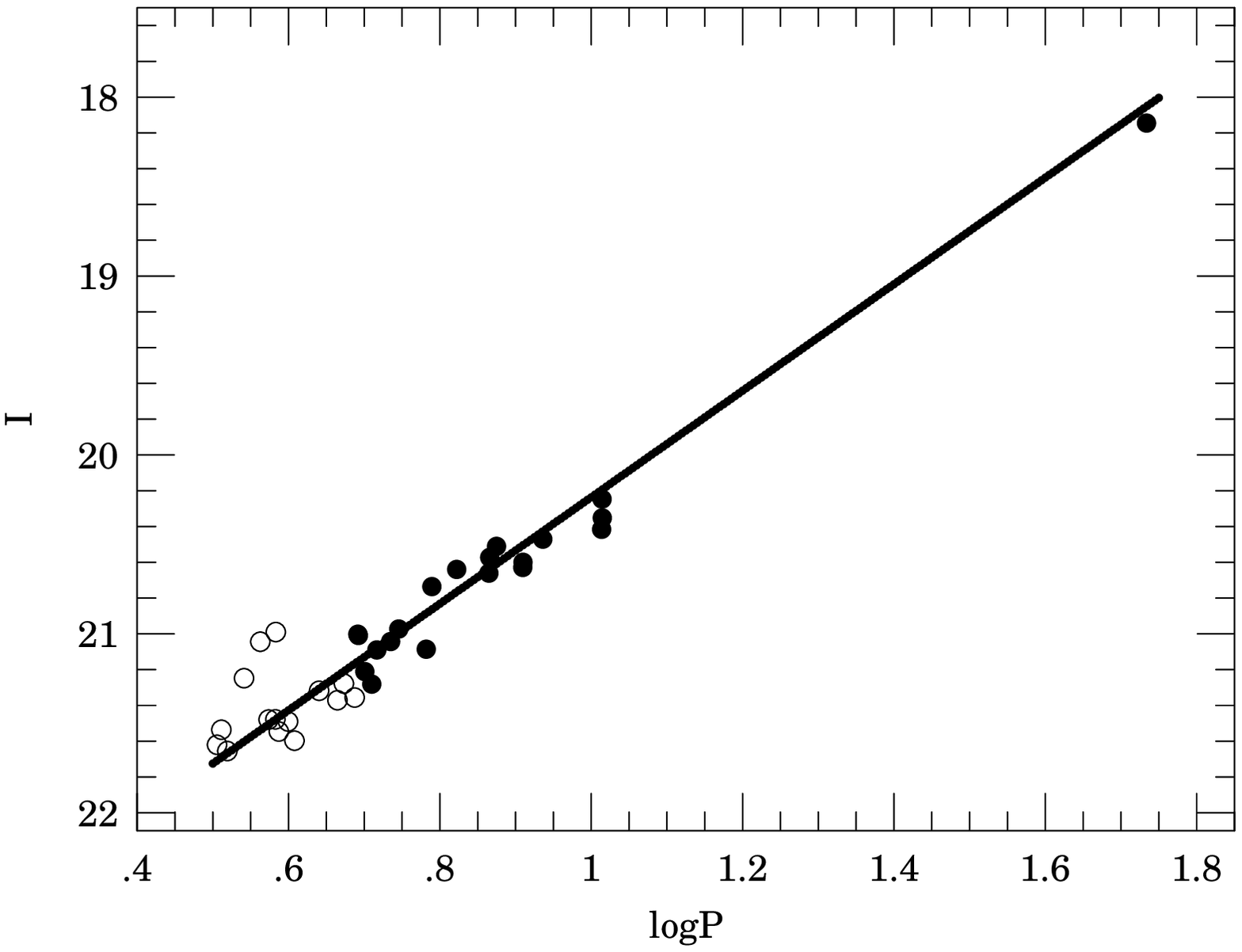}
\caption{Same as Fig. 4, for the I band.
}
\end{figure}

\begin{figure}[htb]
\vspace*{15cm}
\includegraphics{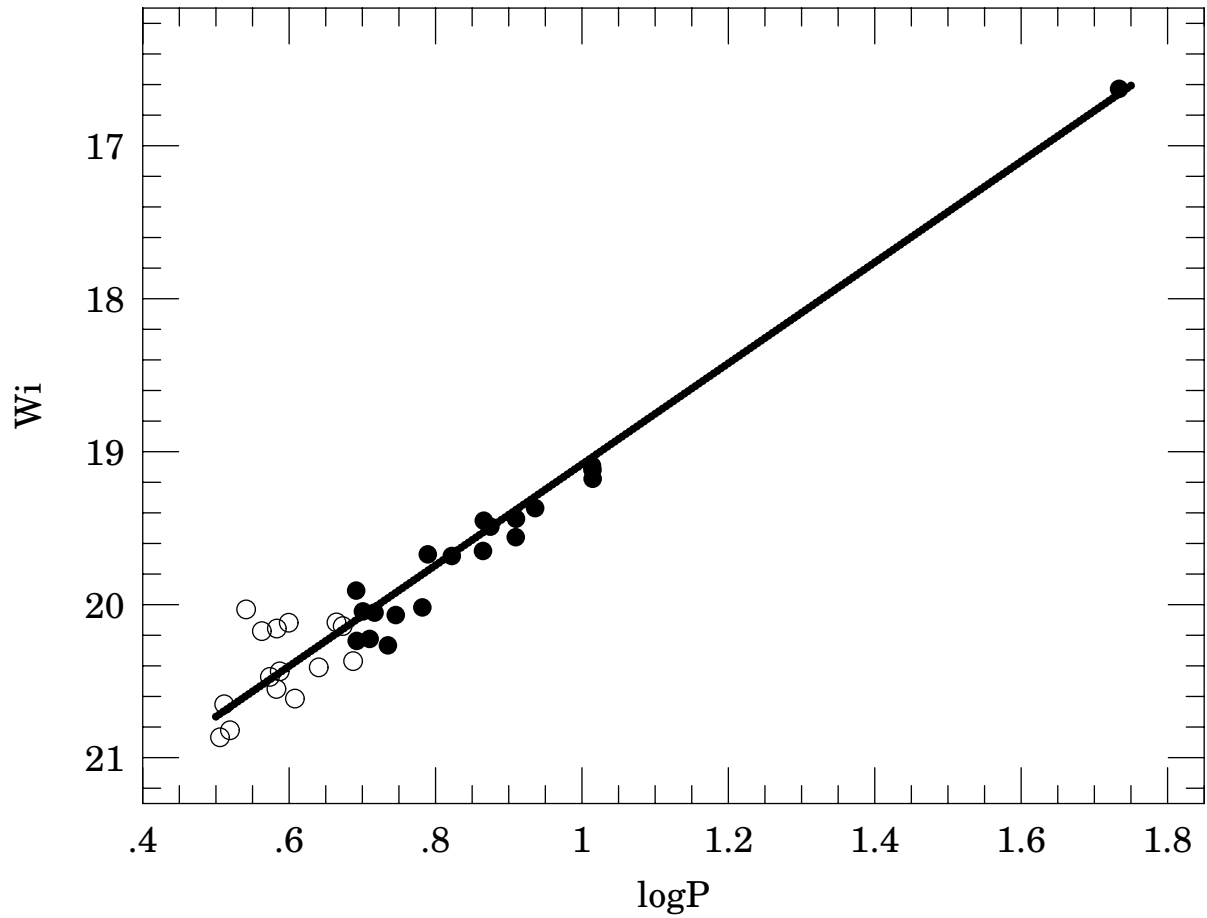}
\caption{Same as Fig. 4, for the reddening-independent (V-I) Wesenheit 
magnitudes. The very small scatter in this diagram for the Cepheids 
denoted by the black circles indicates that this sample is free of
significantly blended stars which would tend to decrease the distance
of WLM derived from these data. 
}
\end{figure}

\end{document}